# Localised surface plasmon resonance spectroscopy: naked nanoparticle sensing

J. L. Spear[1], N. Pliatsikas[2], N. Kalfagiannis[1], P. Patsalas[2], D. C. Koutsogeorgis[1]

[1] School of Science and Technology, Nottingham Trent University, Nottingham, NG11 8NS, UK

[2] Department of Physics, Aristotle University of Thessaloniki, Thessaloniki, GR – 54124, Greece

Please direct all correspondences to J. L. Spear

School of Science and Technology

Nottingham Trent University

Nottingham

NG11 8NS

Email: n0308260@my.ntu.ac.uk

Extension: 83163

**Abstract**

We present a simplified yet sophisticated variation to localised surface plasmon resonance spectroscopy, which makes use of naked or non-functionalised, nanoparticle templates. These nanoparticle templates, produced with a rapid and scalable process, namely laser annealing, were used as a highly sensitive surface sensor to monitor the adsorption of both metallic lead and a lead salt from aqueous solutions, showing a measurable optical response due to a surface abundance of lead as low as 100 ppm from 0.3 ml of $Pb_2SO_4$ solutions, with concentrations less than 20 ppm. This proposed method enables the end user to rapidly assess the surface abundance of lead from a simple optical reflectance measurement and could serve as a platform for *in situ* analysis within water filtration and cleaning systems.

# 1 introduction

In depth knowledge of the localised surface plasmon resonance (LSPR) effect has enabled the development of a diverse range of applications [1-11], such as surface sensing *via* LSPR spectroscopy. LSPR spectroscopy is a label-free and powerful surface sensing platform with sensitivity comparable to other surface sensing techniques, with the added benefits of relatively simple sensor fabrication and measurement equipment, resulting in a surface sensing technique that has gained remarkable popularity [12]. The high quality of LSPR spectroscopy sensors arises from the extreme chemical sensitivity of a plasmonic nanoparticle template (PNT) to minute changes in the local dielectric environment / refractive index, which manifests as a discrete change to their optical response due to surface adsorption of, for example, a heavy metal such as lead [13-18]. The filtration of heavy metals form a water supply is of vital importance, as a build-up of heavy metals in the human body has a severe and negative impact to health, due to the lack of a biological route to dissipate the accumulation of heavy metals. Despite the diverse nature of LSPR spectroscopy, a distinct issue is the need to functionalise the NPT in order to promote maximum and uniform surface adsorption of the analyte [19-21]. While this is in some cases of vital importance, it is an inherently complex fabrication process, especially in the case of duplexed sensors, which acts to limit the scope of a single sensor to the analyte favoured by the functionalization. Here we present an alternative and simplified sensor fabrication technique for LSPR spectroscopy, making use of a naked (non-functionalised) plasmonic nanoparticle template. We demonstrate the ability of such a naked PNT to act as a highly sensitive sensor that can be used to monitor both the surface abundance of lead and the concentration of the lead salt solution the naked PNT was exposed too.

# 2 Materials & Methods

Thin films of Ag were deposited on n-type, (100) Si wafers via radio-frequency magnetron sputtering using: an Ag target (power density of 0.88 $Wcm^{-2}$); and an Ar environment (working pressure of 0.67 Pa). The deposition time was adjusted to produce films with an effective thickness of 10 nm. The thin films were processed with UV LISA (laser induced self-assembly, 1 pulse) and UV MONA-LISA (modification of nanoparticle arrays – laser induced self-assembly, multiple pulses), with a KrF (248 nm) laser system (25 ns pulse length, 1 Hz repetition rate). This process is described in greater detail elsewhere [22]. Four specific laser fluences (350, 500, 650 and 800 $mJcm^{-2}$) and six different treatment pulses (1 – 6 pulses) were used to produce several identical laser processed samples for analysis. Solutions of the lead salt $Pb_2SO_4$ were produced at specific concentrations (0 – 20 ppm, in steps of 4 ppm) in equal parts ethanol and water, due to the hydrophobic nature of the PNT's. The samples were initially treated with 0.1 mL of ethanol to ensure maximum and uniform wetting of the PNT by 0.2 mL of the $Pb_2SO_4$ solutions before evaporating. The optical response of the templates (before and after the evaporation of the $Pb_2SO_4$ solutions) were assessed *via* normal incidence optical reflectance spectroscopy (ORS) with: a white light source; co-axial reflectance probe; and a CCD spectrometer. The surface abundance of $Pb^{++}$ and $PbSO_3$ was identified with high resolution x-ray photoelectron spectroscopy (XPS) measurements with: a monochromatic Al Kα source; hemispherical analyser; and multichannel detector.

## 3 Results & Discussion

Firstly, it was necessary to identify if a naked PNT would provide a measurable spectral shift due to the surface presence of lead. To ascertain the usefulness of the naked PNT's, the optical reflectance was measured before and after the evaporation of the range of specific concentration lead salt solutions. We focus our discussion on two nanoparticle templates: the 500 mJcm$^{-2}$, 1 treatment pulse (1x500) and 500 mJcm$^{-2}$, 6 treatment pulses (6x500) templates.

The spectral shift, Δλ (nm), is defined as the change in spectral position of the primary LSPR reflectance peak due to the surface presence of lead:

$$\Delta\lambda = \lambda_{Pb} - \lambda_{Naked\ NPT} \quad [3.1]$$

Where $\lambda_{Naked\ NPT}$ is the spectral position of the primary LSPR reflectance peak of the naked PNT, prior to exposure to the lead salt solutions, and $\lambda_{Pb}$ is the peak position after the surface adsorption of $Pb^{++}$ and $PbSO_3$ from the lead salt solutions.

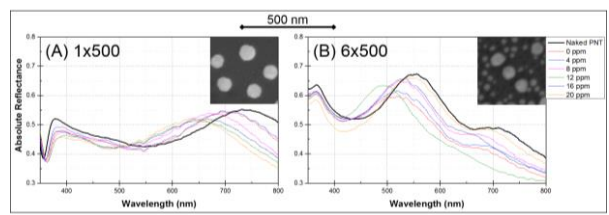

Figure 1: Optical reflectance spectroscopy (ORS) measurements of the naked nanoparticle templates before and after the treatment of the various $Pb_2SO_4$ solutions for the (A) 500 mJcm$^{-2}$, 1 pulse and (B) 500 mJcm$^{-2}$, 6 pulses templates. The associated SEM micrograph inserts show the structure of the as fabricated, naked PNT with the scale bar shown.

As can be seen if Fig. 1, the response of both presented naked PNT's clearly show the surface presence of lead providing a measurable shift of the optical reflectance, however it is evident that any given solution does not induce a uniform shift across both these templates. The 1x500 template shows a clear relationship between the solution concentration and the measured spectral shift (Fig. 1 (A) and the black data points in Fig. 2 (A)), whereby the spectral shift increases with the solution concentration. Conversely, the 6x500 template shows an inverse relationship (Fig. 1 (B) and the black data points in Fig. 2 (B)), where the spectral shift decreases with increasing solution concentration. This suggests that the sensitivity of a naked nanoparticle sensor is dependent on the underlying structure.

To quantitatively relate the measured spectral shift to the surface abundance of lead, the PNT's were further analysed with XPS to identify the surface abundance of $Pb^{++}$ and $PbSO_3$. It can be seen in Fig. 2 (A), that the 1x500 template demonstrates a clear relationship between the solution concentration and the surface abundance of both $Pb^{++}$ and $PbSO_3$, which echoes the link seen previously with the spectral shift. However, the 6x500 template, Fig. 2 (B), does not exhibit such a promising relationship, as the surface abundance of lead can be seen to not show a distinct trend with solution concentration, providing further evidence that the sensitivity of a naked template can be controlled by the nanoparticle structure.

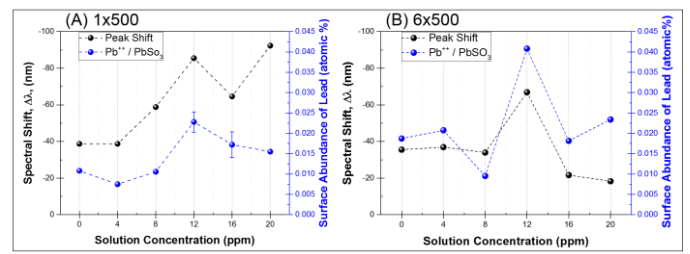

Figure 2: Comparison between the spectral shift (via ORS) and the surface abundance of $Pb^{++}$ and $PbSO_3$ (via XPS) for the (A) 500 mJcm$^{-2}$, 1 pulse and (B) 500 mJcm$^{-2}$, 6 pulses templates. The dashed lines linking the data points are included only as an aid to the eye.

The power of a LSPR spectroscopy sensor is derived from the ability to assess surface adsorption rapidly and with a simple optical measurement. To demonstrate the ability of a naked nanoparticle template to act as a high-quality sensor for detecting and monitoring the surface presence of lead (as both $Pb^{++}$ and $PbSO_3$) a calibration is necessary. A calibration between the measured spectral shift and the surface abundance of lead, as well as between the spectral shift and the original solution concentration was performed for the 1x500 template. This template was selected because of the proportional nature of the relationships it presented.

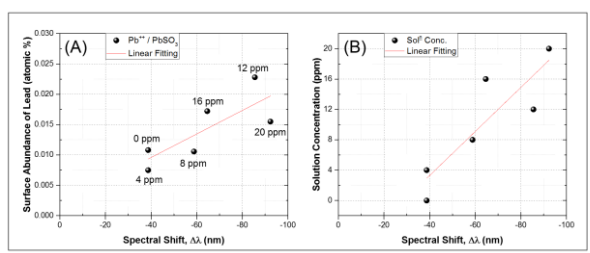

*Figure 3: Calibration curves relating the measured spectral shift (A) the surface abundance of $Pb^{++}$ and $PbSO_3$ and (B) the original solution concentration for the 500 mJcm$^{-2}$, 1 pulse naked nanoparticle template.*

Surface Abundance of Lead (at%) = $-1.9 \times 10^{-4} \cdot \Delta\lambda$ (nm) + $1 \times 10^{-3}$     **[3.2]**

Solution Concentration (ppm) = $-0.29 \cdot \Delta\lambda$ (nm) $-8$     **[3.3]**

The calibration curves, shown in Fig. 3, and subsequently equations [3.2] and [3.3], demonstrate the ability of a naked nanoparticle template to act as a surface sensor for detecting and monitoring the surface abundance of lead, as well as the concentration of the solution the nanoparticles were exposed to.

## 4 Conclusions

In this work, we have presented an alternative, simplified yet sophisticated route towards a highly sensitive LSPR spectroscopy sensor for the detection of the surface adsorption of Pb. While it is traditional within LSPR spectroscopy to make use of functionalization, we provide a demonstration that a lack of functionalization does not equate to a lack of a sensor, as a naked plasmonic nanoparticle template can be used to detect a surface abundance of 100 ppm (0.01 atomic %) of $Pb^{++}$ and $PbSO_3$. The calibration between the spectral shift and the surface abundance of lead and of the concentration of the solution used to treat the naked PNT, enables the user to identify these factors from two simple optical reflectance measurements. Additionally, we have presented that the underlying structure of the nanoparticle template has a significant impact on the inherent sensitivity of the produced sensor.

**Acknowledgements:** J.L. Spear would like to acknowledge funding under the Vice Chancellor PhD Bursary scheme of Nottingham Trent University. The authors would like to acknowledge Prof. Wayne Cranton and Stuart Creasey of Sheffield Hallam University, Sheffield, UK for providing training and access to the FEG-SEM used for this work.